\documentclass[prb,apsrev,floatfix,onecolumn]{revtex4-1}

\usepackage{amsmath}
\usepackage{graphicx,color}
\usepackage{amsfonts}
\usepackage{amssymb}
\usepackage{graphicx}%
\providecommand{\U}[1]{\protect\rule{.1in}{.1in}}

\begin{document}

\title{
%
%
%
Role of disorder and fluctuation on charge migration dynamics in molecular aggregate with quantum mechanical network }

\date{\today}
\author{Takehiro Yonehara}
\email{tkyn2011@gmail.com}
\affiliation{RIKEN Center for Computational Science, Kobe 650-0047, Japan}

\begin{abstract}
We examine the effect of structural disorder and dynamical lattice fluctuation 
on charge migration dynamics starting from a birth of local exciton 
in a quantum network of molecular aggregates 
by using model Hamiltonians having complicate interactions. 
Here all monomers are supposed to be the same for simplicity.
A natural use of inherent sparsity of Hamiltonian matrix allows us an investigation of 
essential features in quantum network dynamics accompanied with a fluctuation of interaction. 
Variation of disorder parameter, kinetic energy and effective mass of monomers   
in electron dynamics calculation reveal how static disorder and dynamical fluctuation affects 
electron dynamics in a large size of molecular aggregates. 
Disorder in aggregate structure suppress charge separation 
while molecular motion can promote charge diffusion in cases of smaller mass.
These findings are obtained by using a newly introduced formula 
for evaluating charge separation in molecular aggregates  
that is useful for other analysis involved with charge migration dynamics  
in molecular/atom aggregates. 
This work provides a way for obtaining a quantum mechanical time-dependent picture of 
diffusion and migration of exciton and charge density in general molecular aggregates, 
which offers a fundamental understanding of electronic functionality of nanocomposites. 
\end{abstract}

\maketitle

\tableofcontents



%
%
%
\section{Introduction}

Exciton and following charge migration is important and ubiquitous 
in photo energy conversion process in material.  
A photo excited exciton is a source of charge separation leading to chemical driving force.
\cite{Sundstrom-review-solar-energy,accum-CS}
Fundamentally and microscopically, they are described by quantum dynamics.
%
%
%
Though a picture of charge separation is still unrevealed, 
experimental observations of time dependent charge separation and 
their control have been reported in pico second and nano second time scales.
\cite{Nakajima-JPCL2012-CS-obs-ps,image-CS-CR}
In order to understand charge separation mechanism  
we need a time dependent picture fine enough to include chemical reaction 
within sub-pico-second.

Because a fully ab initio description of charge separation in quantum dynamics theory 
is computationally tough, 
an investigation of interesting features related to 
a realistic size of molecular aggregate system 
accompanied with fluctuation have been limited. 
To overcome this, we utilize a simplified tight-binding model 
having sparsity in a molecular aggregate Hamiltonian matrix, 
referential occupation of quantum sub-states inside monomers 
and Liouville--von Neumman equation.
We can reduce a computational cost using this sparsity of 
interaction network in molecular aggregates 
associated fast spatial damping of electron coupling expected in realistic material system. 

Based on this, we demonstrate an efficient scheme 
for examining a time dependent behavior of charge separation. 
We investigate how static and dynamical disorder affect a charge separation 
after a birth of exciton in molecular aggregate system.  
Here, static and dynamical disorders mean 
initial disorder in reference structure and 
time-dependent deformation of configuration of aggregate, respectively.

Here we examine a quantum dynamics of excited electrons 
without resorting to method using too much coarse graining treatment 
such as kinetic Monte Carlo schemes 
because they can never tell about a microscopic origin 
inherent in this kind of phenomena, for example, 
quantum mechanical interference, decoherence and dynamical resonance, and so on. 
These methods are complementary. In fact, the latter Monte Carlo type kinematic methods 
have been successfully used for the description in a wide range 
from middle to macroscopic scale. 
\cite{Madigan-PRL2006,Sousa-JCP2018}
%

%
%
We here focus on a shorter time scale and smaller systems compared to such a macroscopic view, 
namely, middle scale description with respect to time and spatial size. 
By introducing and using a new concise scheme, 
we intend to propose an origin of excitation driven charge migration 
in a moderate size of molecular aggregate systems.

In the next section, \ref{theory}, 
theoretical method is explained by focusing 
a Hamiltonian sparsity for molecular aggregates.  
That is followed by the numerical section, \ref{numerical}, 
for demonstrating the present scheme 
using fused lattice structure system consisting of 
simplified monomers having local quantum sites. 
There, we introduce charge diffusion and separation measure and 
discuss effects of structure fluctuation and static disorder. 
In Sect. \ref{concl} we conclude this article with a future perspective
about the present scheme.

\section{Theoretical method}
\label{theory}


We briefly explain the general formulation of the developed calculation method 
aimed for a concise description and examination of exciton and charge migration. 
Generalization with respect to state number and couplings is formally straightforward. 
The sparsity of Hamiltonian matrix is utilized for attaining a reduction of computational cost.

\subsection{Sparse form of Hamiltonian matrix
having a network structure in interactions of quantum states}

For a compact and economical calculation of quantum dynamics 
supported by a sparcity in a network associated with Hamiltonian matrix, 
instead of treating full matrix elements  
\begin{align}
\{ H_{\alpha \beta} \}_{ \alpha, \beta = 1 \sim N_{\rm{basis}} }, 
\label{H_orginal}
\end{align}
we employ a following data structure as the expression of this matrix, 
\begin{align}
\{ \alpha_i,  \beta_i, H_{\alpha_i \beta_i} \}_{ i = 1 \sim N_{\textrm{pair}} }.
\label{H_sparse}
\end{align}
Here $N_{\textrm{pair}}$ is a number of non-zero elements in the original matrix. 
To be more precise practically, for an infinitesimal threshold value $\eta$, 
$N_{\textrm{pair}}$ is the number of elements in the set 
$ \{ H_{\alpha \beta} ; | H_{\alpha \beta} | \ge \eta \} $. Practically 
we can establish $\eta$ as a function of a spatial distance of pair subsystems. 

Next, we provide a block structure associated with a molecular aggregation 
that we are interested in here.
Total basis number $N_{\textrm{basis}}$ is the same as 
a sum of numbers of quantum states given for all subsystems,  
$ N_{\textrm{basis}} = \sum_{a=1}^{N_{\textrm{g}}} M_I(a) $,
with $ M_I(a) $ being the number of internal quantum states in subsystem $a$. 
$N_{\textrm{g}}$ is a number of group sites. 
The total Hamiltonian matrix is divided into $N_{\rm{g}}^2$ sectors and 
the $(a,b)$-th block takes a matrix form having a size of $M_I(a) \times M_I(b)$ 
where $a$ and $b$ in $(a,b)$ denote positions of row and column, respectively.
$ a = b $ and $ a \neq b $ correspond respectively to 
local Hamiltonian of subsystem $a$ and coupling matrix between $a$ and $b$. 
Practically, the number of multiplication of their elements 
to other matrix is reduced effectively 
accompanied with a small additional cost 
in pointer skip in a memory space. 
We refer to this as a compressed Hamiltonian matrix (CHM).

As a final part of this subsection, for making clear the sector-ed structure,  
we define a one-to-one mapping 
between an index label of original matrix and an index pair in sector-ed matrix 
as $\{ k \Leftrightarrow (a,\alpha_a) \} $
with 
$  1  \leq  k  \leq  N_{\textrm{basis}} $,  
$  1  \leq  a  \leq  N_{\textrm{g}}   $  
and $ 1  \leq  \alpha_a  \leq  M_I(a)  $.

\subsection{Liouville--von Neuman equation 
using multiplication of compressed Hamiltonian matrix}
Using CHM, 
a time derivative of density matrix $\rho_{\alpha\beta}$ in LvN equation, 
\begin{align}
\dfrac{d\rho_{\alpha\beta}}{dt}
=
- \dfrac{i}{\hbar}
\sum_{\gamma}^{Nb}
\left\{
\rho_{\alpha\gamma} H_{\gamma\beta} - H_{\alpha\gamma} \rho_{\gamma\beta} 
\right\}, 
\end{align}
can be cast into the following pseudo code: 
\begin{align}
&\omega  := 0   \notag \\
&do \quad i = 1, \,\, N_{\textrm{pa}} \notag \\
&\quad do \quad  p = 1, \,\, N_b  \notag \\
&\quad\quad \omega_{p\beta_i} := \omega_{p\beta_i} -
\dfrac{i}{\hbar}  \rho_{p\alpha_i}  H_{\alpha_i\beta_i} \notag \\
&\quad enddo \notag \\
&\quad do \quad  q = 1, \,\, N_b  \notag \\
&\quad\quad \omega_{\alpha_i q} := \omega_{\alpha_i q} -
\left( - \dfrac{i}{\hbar}   H_{\alpha_i\beta_i} \rho_{\beta_i q}   \right)   \notag \\
&\quad enddo \notag \\
&enddo 
\label{LvN-sparse}
\end{align}
with 
$ \omega \equiv \dfrac{d}{dt} \rho $.

\subsection{Structure of model Hamiltonian employed here}

We briefly summarize a specific form of Hamiltonian 
employed here for a demonstration of the present method.

Each block bare sub-Hamiltonian sector for subsystem a is 
$ H^{aa;0}_{i_a j_a} = \delta_{i_a j_a} \epsilon^{a}_{i_a} $.
Coupling off-diagonal block sectors for different subsystems, a and b, are given by 
$ 
H_{i_a j_b}^{ab}(R_{ab}) = 
s_{i_a i_b}^{ab} \rm{exp}\left( - R_{ab} / R^{\rm damp}_{i_a j_b} \right) 
$. 
Here, $R_{ab}$ denotes distance between subsystems a and b 
while $ R^{\rm damp}_{i_a j_b} $ are the parameters that 
determine a damping strength with respect to $R_{ab}$. 
Range of indicies are $ a, b = 1, 2, ..., N_{\textrm{g}} $ and $ i_a = 1, 2, ..., M_I(a) $.
For taking into account an effect of structure dynamics in a simulation,  
diagonal elements in each block sub-Hamiltonian sector for subsystem $a$ were designed to change 
as a function of distance of each pair subsystems 
in a form of 
$  H_{i_a i_a}^{aa} =  H^{aa;0}_{i_a i_a} 
   + \sum_{b \neq a} \dfrac{1}{4} k^\textrm{spring} (R_{ab}-R_{ab}^\textrm{ref})^{2} $.
In this equation we used 1/4 but not 1/2 for avoiding an apparent double counting  
so that the spring constant, $k^\textrm{spring}$, should keep physical meaning.

%

\subsection{Time propagation}

For computational efficiency, 
we use the Chebyshev expansion method for the time propagation operator 
associated with LvN equation of motion. 
The formalism of this time propagation method is summarized 
in Appendix \ref{Cheby-LvN} for self-containednes, 
of which theoretical details are given in the work by Guo {\it et al.} 
\cite{Guo-JCP1999-Cheby-LvN}

The time increment for each time step used in this article was set to 4 a.u., 
which was sufficient for the conversion of the result, now shown here.   
The number of time steps in simulation is 500 and total time is 2000 a.u. 
which is 48.3 fs.

\section{Numerical demonstration}
\label{numerical}

\subsection{Effective Charge distance from reference point}

Here we examine how a charge pattern emerges 
after a birth of local exciton at the center 
in a generally disordered lattice system 
consisting of molecular monomers. 

This preparation of initial excitation is intended 
for examining inherent feature of dispersion 
that a quantum mechanical network has. 

For this purpose, we construct an evaluation formula 
for effective position of wave fronts represented by 
$u^+$ and $u^-$ for positive and negative charge 
roughly measured from position of initial exciton.

\begin{itemize}
\item[1)]
Calculate sum of positive and negative charge over the sites as \\
\begin{align}
q_\textrm{sum}^{\pm} = \sum_{g}^\textrm{site} h(\pm q_g) \, q_g \quad 
\end{align}
$h(x)$ is the Heaviside function and takes 1 and 0 respectively for x $>$ 0  and x $\le$ 0. 
\item[2)]
Evaluate weighted sum of positive negative charge multiplied with 
the distance from the reference position over the sites as \\
\begin{align}
u^{\pm} = \sum_{g}^\textrm{site} h(\pm q_g) \, q_g |{\bf r}_g-{\bf r}_\textrm{ref}| \quad  
\end{align}
\item[3)]
Normalize $u^+$  and $u^-$ by using $q_\textrm{sum}^+$ and $q_\textrm{sum}^-$.   
And then, determine the effective propagation front of positive and negative charge as 
\begin{align}
 d^{\pm} = u^{\pm} / q_\textrm{sum}^{\pm}
\end{align}
\end{itemize}

The charge on each site is evaluated as 
the difference of quantum population in the corresponding site 
from the reference occupancy. 
In the following numerical demonstration,  
each site has two quantum states of local ground and excited states 
with $ M_I = 2 $  and an reference occupation only in the ground state. 

For each initial structure we prepared randomly 30 velocity vectors 
for the geometries having disorder parameter $f$ being fixed  
and took average and variations of $ d^{\pm} $. 

Magnitude of each velocity vector of monomer was 
determined so as to match with equivalently divided amount 
of energies according to micro-canonical temperature, 0 and 300 K  
while the directions of vectors were given randomly.

\subsection{Details}

 First of all in the presentation, we summarize the calculation conditions 
including system parameters employed here. 

 For simplicity we set the two internal quantum states for each, namely, 
we set $ M_I(a) = 2 $ with a referential occupancy for lowest one 
be hypothetical setting of use of HOMO and LUMO for arbitrary monomer $a$ in the system.
Below we express $ M_I(a) = M_I $. 
Since a three dimensional calculation is computationally demanding, 
we treat two dimensional model that still has a significance as 
a starting point for investigating electron dynamics in a surface of molecular crystal. 
The energies of these two states are commonly set to be 
$\epsilon^a_{1}=0.0$ and $\epsilon^a_{2}=0.3$ Hartree for all $a$ 
that appear in each block matrix placed at diagonal position 
in a whole Hamiltonian matrix. 
We selected this energy difference $\epsilon^a_{2}-\epsilon^a_{1}=0.3$ 
similar to HOMO-LUMO gap of naphthalene obtained by CAM-B3LYP/6-31G(d) calculation 
at optimized geometry, 0.2731 Hartree. 
We note that a naphthalene serves as an electron donor molecule,  
for example, in a naphthalene - tetra-cyano-ethylene dimer. 
Therefore, the employed test model of molecular aggregate 
is expected to have a hole transfer property. 

As a referential spatial configuration of monomer positions, 
we employed square lattice on x-y plane 
of which numbers of rows and columns are expressed by $n_x$ and $n_y$ with $n_x=n_y$. 
The total number of monomers are $ N_{\textrm{g}} = n_x n_y $.
In this article, we examine two cases of $n_x=n_y=25$ and $31$
with fixed reference distance of nearest monomers being $ D_{x} = D_{y} = 7.8$, respectively. 
Later we simply express $D_{x}=D_{y}$ as $D$. 
Note that in dynamics calculations they change 
according to the introduced disorder of geometrical structure at initial simulation time 
and time-dependent deformation of configuration, 
which is followed by a clear illustration of a significant effect 
on charge migration dynamics after a birth of local excitation. 
Total number of elements in a basis set are given by 
$ N_{\textrm{basis}} = N_{\textrm{g}} M_I $.


We introduced a disorder to initial geometry of locations of monomers in aggregate 
by giving random displacement $ ( \eta -0.5 ) f D $ for them
in each Cartesian coordinate, x and y 
compared from reference positions within a lattice structure.  
Here $ \eta $ being random number over $[0:1]$ while 
$f$ is a parameter that modulate the degree of disorder, that we call disorder parameter. 
In this article, we compare the results with variation of $f$ = 0.0, 0.2 and 0.4.   

Time increments in dynamics calculation commonly used for quantum and classical 
degrees of freedoms are commonly set to efficiently and practically small value, 4 au. 
Total simulation time is $T_{\textrm{max}} = 2000$ 
with a number of time steps $N_{\textrm{t}}$ being 500.

Then, we briefly explain model functions used in construction of Hamiltonian matrix.
The elements of damping matrix of monomer interaction matrices $H_{i_a i_b}^{ab}(R_{ab})$ 
were set as 
$R^{\textrm{damp}}_{i_a j_a }=1.2$ \AA \,( = 2.268 Bohr )
for all $i_a,j_a=1 \sim M_I(a)$ and $a$
while strength amplitude matrices in integration matrices 
were supposed to be 
$s_{11}^{ab}=0.3$, $s_{12}^{ab}=s_{21}^{ab}=1.5$, and $s_{22}^{ab}=1.0$. 
A threshold parameter for monomer pairing interaction is set to $r_{\textrm{pair}}$ = 10.8. 
Monomer interactions in a total Hamiltonian matrix were taken into account 
only for monomer pairs having distances being less that $r_{\textrm{pair}}$, 
which characterize sparsity of Hamiltonian for aggregate system. 
We checked that this value of $r_{\textrm{pair}}$ is sufficiently large 
for the convergence of results with use of 
damping matrix $R^{\textrm{damp}}_{i_a j_a }$ and 
strength matrix $s_{i_a j_b}^{ab}$ in monomer interaction mentioned above.

Masses of uniform monomer were varied as 
$\mu = 5\times 10^4$, $1 \times 10^5$ and $2 \times 10^5$
for examining a kinematic effect on charge migration dynamics. 
The spring constant of hypothetical harmonic potential for each monomer pair 
was commonly set to $k^{\textrm{spring}}=0.002$ throughout dynamics calculations. 
In order to obtain a fundamental information on a propensity of charge migration dynamics,  
we utilized an introduction of a local exciton as a sudden perturbation 
in quantum state of whole the system at the initial simulation time. 
In all the dynamics calculations, 
central monomer was initially locally excited from the lowest to highest state 
with a half amount of quantum population,
which corresponds to hypothetical HOMO-LUMO excitation.

Each portion of initial kinetic energy in 2$N_{\textrm{g}}$ all the classical degrees of freedom 
of molecular aggregates confined in x-y plane was set to 
$\frac{1}{2}k_{\textrm{B}}T$ with the variation of 
micro-canonical temperature $T$ being 0 and 300 Kelvin.
Here $k_{\textrm{B}}$ denotes the Boltzmann constant. 
Direction of velocity vectors for each monomer was randomly given.
In all the simulations, 
we employed 30 sets of sample initial velocities for taking average of quantum properties. 
In the present simulation time scale,  
effect of a bounce of excitons and front of charge wave 
from peripheral edge of the whole lattice is negligible. 

Finally, we briefly show the efficiency of the present method using Eq.(\ref{LvN-sparse})
based on the sparsity of Hamiltonian of molecular aggregate. 
Comparisons of computational cost with the original Hamiltonian without use of sparsity
are summarized in Tab. \ref{speedup} associated with the variation of 
system size $n_x=n_y$ and $r_{\textrm{pair}}$. 
In the case of $n_x=n_y=31$ having largest computational cost in this article, 
we can obtain about 20 times speed-up compared to the fully connected Hamiltonian.

%
%
%
%
\subsection{Results and discussions}

\begin{figure}[th]
\includegraphics[width=1.0\textwidth]{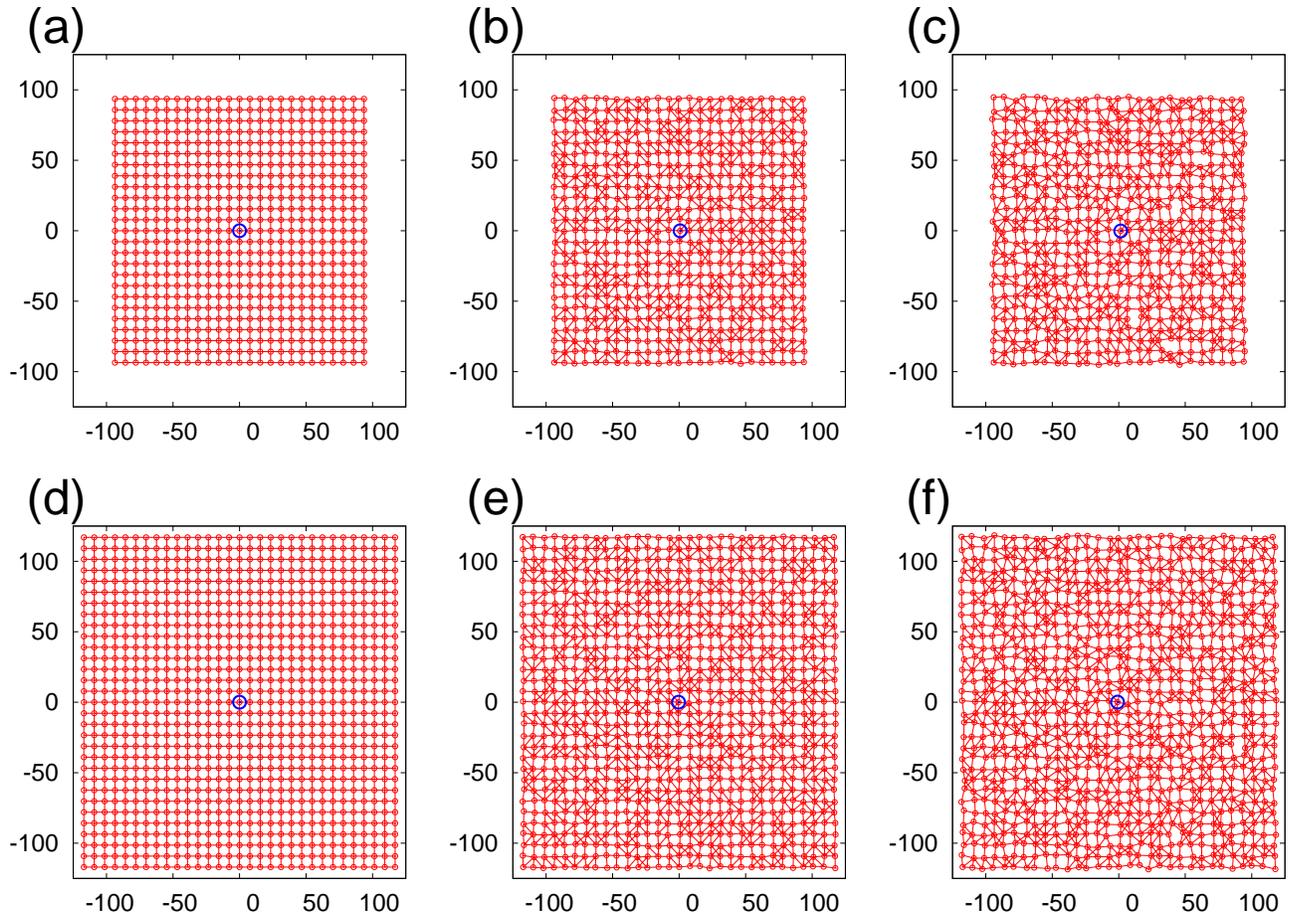}
\caption{
Network structures of aggregates for $n_x$=$n_y$=25 (a,b,c) and 31(d,e,f) cases. 
Disorder parameters $f$ are 0.0, 0.2 and 0.4 respectively for panel (a/d), (b/e) and (c/f). 
Lattice size as reference length of nearest site is $D_{x}=D_{y}=7.8$ Bohr.
$f$ serves as a ratio of fluctuation compared to this reference length.
Rings at the centers of panels denote the site with initial excitation.}
\label{Fig-1}
\end{figure}

\begin{figure}[th]
\includegraphics[width=1.0\textwidth]{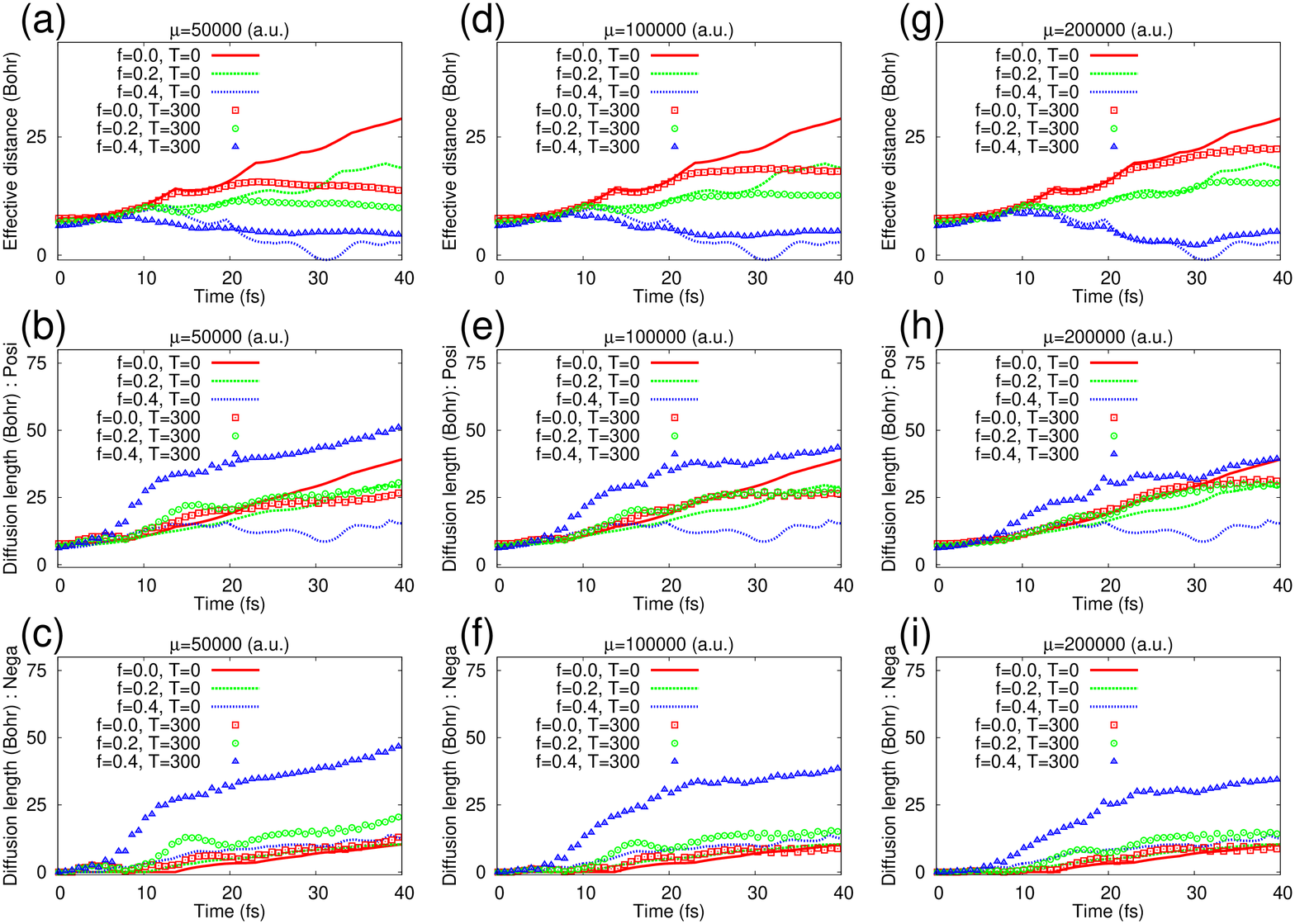}
\caption{
Effect of disorder, mass and structure dynamics on charge migration dynamics. 
Top(a,d,g), middle(b,e,h) and bottom(c,f,i) panels show time dependent behaviors 
correspondingly of effective distance of opposite sign charges,$(d^{+}-d^{-})$ 
diffusion length of positive, $d^{+}$, and negative ones, $d^{-}$.
Masses employed in panels in the left(a,b,c), middle(d,e,f) and right(g,h,i) columns 
are $ \mu = 5\times{10}^4$, $1\times{10}^5$ and $2\times{10}^5$, respectively. 
$n_x$=$n_y$=25.
}
\label{Fig-2}
\end{figure}

   As a general tendency of dynamics independent of $\mu$, $f$ and $n_x=n_y$, 
   the time dependent distance of effective positive and negative charge fronts 
   are reduced by molecular motion 
   that we can see by the comparison between $T=0$ and 300 
   in the panels (a), (d) and (g) of Fig. \ref{Fig-1} and \ref{Fig-2}  
   despite of slight exceptions after the half of simulation time in 
   the case of $n_x=n_y=25$ with $T=300$. 

   In the cases of $n_x=n_y=25$,
   the effect of structure dynamics on charge front propagation  
   critically depends on the disorder parameter $f$.   
   The larger disorder leads to kinetic promotion of charge front propagation 
   as found in the (b/c), (e/f) and (h/i) of Fig. \ref{Fig-1}.  
   To explain more precisely, as seen in the comparison between 
   the results of $(f,T)=(0.4,300)$ and $(0.4,0)$ presented in the panel (b) 
   associated with the smallest mass $\mu=50000$, 
   kinetic motion promotes the propagation of positive charge front. 
   On the other hand, as seen through the comparisons of the cases 
   of $(f,T)=(0.0,300)$ and $(0.0,0)$  in the panel (c), 
   the final position of positive charge front is reduced due to kinetic motion.
   It is also the case tor the propagation of negative charge front 
   that structure dynamics provides larger effect on results 
   with higher disorder parameter $f$.  
   These trends are also seen in the cases of different masses 
   presented in panels (e/f) and (h/i) 
   though the larger mass is accompanied with weaker dependency 
   due to the reduction of weaker modification of monomer coupling during dynamics.

   The regularity in structure characterized by small disorder parameter $f=0$ 
   contributes to charge separation    as found in the time dependent behavior 
   of effective distance of positive and negative charge.
   The kinetic fluctuation with $T=300$ suppress significantly a charge separation 
   and moderately depress both propagation of charge fronts. 

   On the other hand, compared to the simulations without disorder, 
   in case of no kinetic fluctuation expressed by $T=0$, 
   the large disorder associated with $f=0.2$ and $0.4$ suppresses the propagation of wave front 
   of positive and negative charge and resultantly the charge separation is depressed. 
   However, interestingly, in the high disorder cases, 
   the kinetic fluctuation labeled with $T=300$ promotes the propagation of wave front   
   with keeping the small difference of them, namely, depressed charge separation. 

   These tendencies are common to the cases with different lattice sizes.

   The effect of lattice size appears in a case of small mass with $\mu=50000$.
   The case of larger system $n_x=n_y=31$ with the same unit lattice distance  
   is characterized by a progression of the wave front of both sign charges.
   This can be attributed to the coupling of electrons with 
   collective motion of structure.

\onecolumngrid

\clearpage

\clearpage

\clearpage

\begin{figure}[th]
\includegraphics[width=1.0\textwidth]{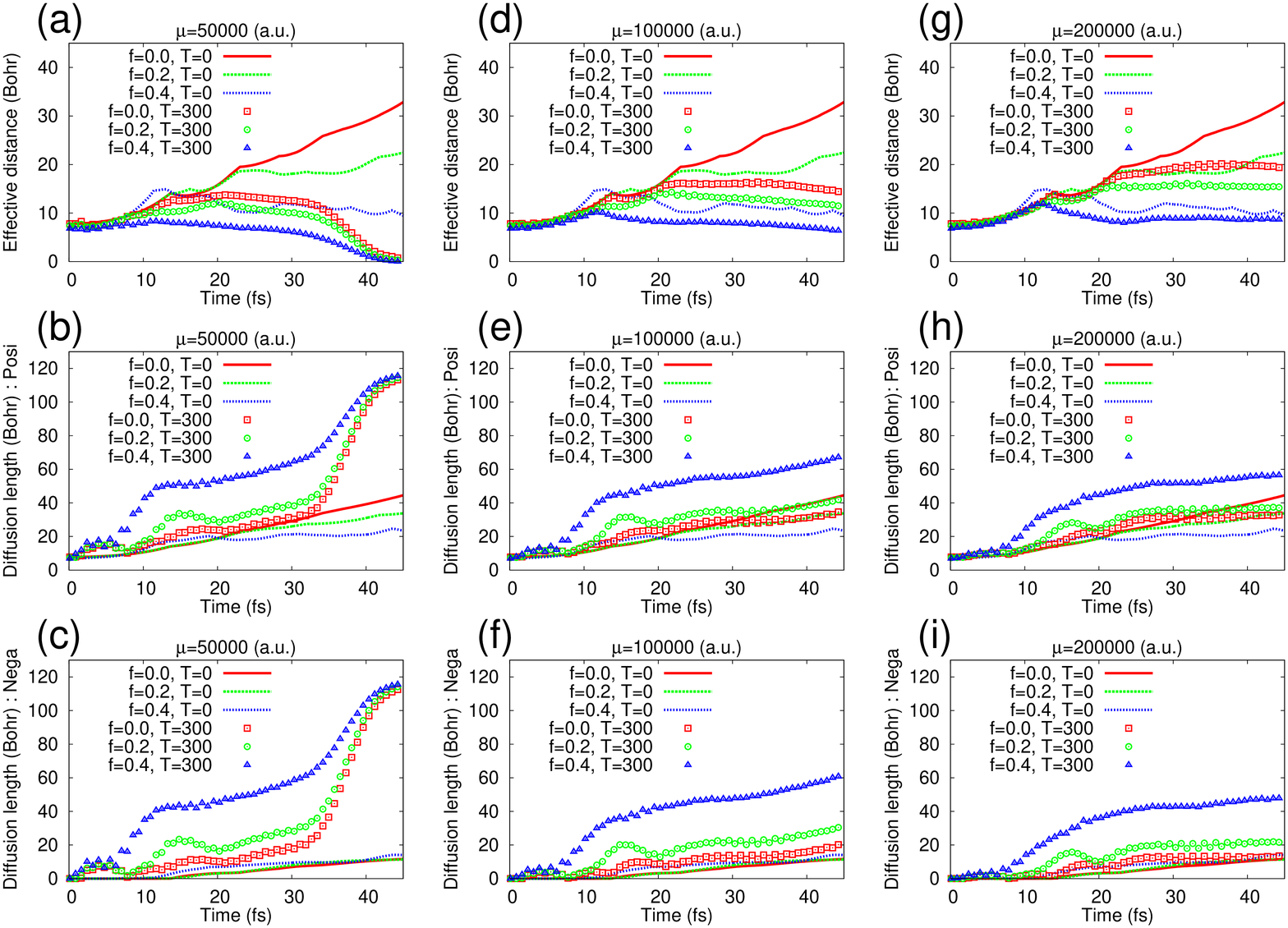}
\caption{
The counterpart of Fig.\ref{Fig-2} for $n_x$=$n_y$=31.
As seen in the cases of small mass here and comparisons with small size cluster, 
kinematic motion affect enhancement of charge propagation 
}
\label{Fig-3}
\end{figure}

\clearpage

%
%
%
\section{Concluding remarks}
\label{concl}
In this article, we numerically clarified the effect of 
structural disorder and dynamical fluctuation 
on charge migration dynamics starting from a birth of local exciton in 
a quantum network of molecular aggregates.
For the convenient analysis, we utilized model Hamiltonians 
having complicate interactions of which parameters were 
determined by using quantum chemical calculation.
Effects of static disorder and dynamical fluctuation on charge migration 
were examined by varying disorder parameter, kinetic energy and effective mass of monomers.  
A summary of observation in the numerical part of this work is as follows: 
(i)  structural disorder in aggregate reduces the rate of charge separation 
(ii) molecular motion can promote a charge diffusion in cases of smaller mass of each monomer.
We can expect that these knowledge support a future realization of 
optimal condition for aimed properties of electronic device 
with respect to a charge separation leading to current in solar cell, 
photo catalysis reaction involved with non-local radical creation, 
and spontaneous photoemission caused by annihilation of electron and hole pair.

In our future work, ingredients of Hamiltonian matrix of aggregates 
are obtained by using, for example, the group diabatic Fock scheme 
\cite{gdf-eld}
combined with locally projection of active orbital space, 
consideration of non-linearity with respect to density matrix 
and spin-orbit interaction.  
\cite{rt-pgdf-eld}
This provides a way for exploring mechanisms 
underlying diffusion and migration dynamics 
of excition and charge density from a quantum dynamics view point, 
which supports a development of electronic functional material.

%
%
%

\begin{acknowledgements} 
This research was supported by MEXT, Japan, 
``Next-Generation Supercomputer Project'' 
(the K computer project) 
and ``Priority Issue on Post-K Computer'' 
(Development of new fundamental technologies 
for high-efficiency energy creation, conversion/storage and use).
Some of the computations in the present study were performed 
using the Research Center for Computational Science, Okazaki, Japan,  
and also HOKUSAI system in RIKEN, Wako, Japan.  
The author deeply appreciates Dr. Takahito Nakajima in RIKEN-CCS 
for the financial support and research environment.
\end{acknowledgements}

\small
\begin{center}
{\bf{AVAILABILITY OF DATA}}
\end{center}
\normalsize

The data that support the findings of this study are available on request from the corresponding author.

\clearpage

%
%

\noindent
\appendix

\section{Chebyshev propagator for LVN equation}
\label{Cheby-LvN}

We here briefly summarize the treatment of time propagation 
using the Liouville--von Neumann equation of motion for a density matrix,  
\begin{align}
\dfrac{\partial \hat{\rho}(t)}{\partial t}
= \dfrac{1}{i\hbar}
\left[\hat{H},\hat{\rho}(t) \right]
\equiv
\dfrac{1}{i\hbar}
\hat{\hat{\mathcal{L}}}
\hat{\rho}(t)
\label{LvN-eq}
\end{align}
with 
$
\hat{\hat{\mathcal{L}}}
\equiv
\left[ \hat{H},\hat{\rho}(t) \right]
$.

 An infinitesimal time propagation operator has a form of 
\begin{align}
\hat{\rho}(t+\Delta t)
=
e^{ - i \hat{\hat{\mathcal{L}}} \Delta t / \hbar }
\hat{\rho}(t)
\label{LvN-eq}
\end{align}
Below we set $\hbar=1$.

%
%
%
 The Chebyshev expansion of time propagator offering 
numerical stability and near uniform accuracy in the entire spectral range  
\cite{Guo-JCP1999-Cheby-LvN}
is expressed by 
\begin{align}
e^{ - i \hat{\hat{\mathcal{L}}} \Delta t }
& =
e^{ - i  L^{+} \Delta t }
e^{ - i \hat{\hat{\mathcal{\widetilde{L}}}} L^{-} \Delta t } \notag\\
&
\approx
e^{ - i  L^{+} \Delta t }
\sum_{k=0}^{K}
( 2 -\delta_{k0} ) 
I_k ( L^{-} \Delta t )
\cdot
\hat{\hat{T_k}}
\left( \hat{\hat{\mathcal{\widetilde{L}}}} \right), 
\end{align}
with 
$ L^{\pm} \equiv ( L_{\textrm{max}} \pm L_{\textrm{min}} ) / 2 $ , 
$
\hat{\hat{\mathcal{\widetilde{L}}}}
\equiv 
( - \hat{\hat{\mathcal{L}}} + \hat{\hat{I}} L^{+} ) / L^{-}$ ,
$  \hat{\hat{T_k}} ( \hat{\hat{\mathcal{L}}} )  =
   \textrm{cos}( k \cdot \textrm{arccos} ( \hat{\hat{\mathcal{L}}} ) ) $.
$L_{\textrm{max}}$ and $L_{\textrm{min}}$ are maximum and minimum eigen values. 
$I_k(x)$ is the k-th order first-kind modified Bessel function 
defined by $ I_k(x) \equiv i^{-k} J_k(ix) $ with 
the k-th order first-kind Bessel function $J(x)$. 
The form of $J_k(x)$ extended from a real value variable $x$ 
to complex plane $z$ is expressed by 
$ J_k(z) = \dfrac{1}{2 \pi i} \oint e^{(z/2)(t-t^{-1})} t^{-n-1} dt $. 
$J_k(x)$ can also be defined through the Laurent expansion of $e^{(z/2)(t-t^{-1})}$ by 
$e^{(z/2)(t-t^{-1})} = \sum_{k=-\infty}^{\infty} J_k(z) t^k$.
This also has the relation of $ J_{-k}(x) = (-1)^k J_k(x) $.

Thus, the Chebyshev expansion of density operator propagated from the previous time 
can be written in a practical form as 
\begin{align}
\hat{\rho}(t+\Delta t)
\approx 
e^{ - i  L^{+} \Delta t }
\sum_{k=0}^{K}
( 2 -\delta_{k0} ) 
I_k ( L^{-} \Delta t )
\hat{\rho}_k,
\end{align}
where,
\begin{align}
& \hat{\rho}_k = 
2 \hat{\hat{\mathcal{\widetilde{L}}}} \hat{\rho}_{k-1}
- \hat{\rho}_{k-2} \quad ( k \ge 2 ) \quad \textrm{and} \\
&\rho_0 \equiv \rho(t), \quad 
\rho_1 \equiv \hat{\hat{\mathcal{\widetilde{L}}}} \rho(t).  
\end{align}
Note that 
\begin{align}
\hat{\hat{\mathcal{\widetilde{L}}}} \hat{\rho}
=
\left[ \hat{\widetilde{H}} , \hat{\rho}  \right]
+
\dfrac{L^{+}}{L^{-}} \hat{\rho}
\end{align}
with 
\begin{align}
\hat{\widetilde{H}}  \equiv
\left( - \hat{H} + \hat{I} L^{+} \right) / L^{-}
\end{align}
since 
\begin{align}
&
\hat{\hat{\mathcal{\widetilde{L}}}} \hat{\rho}  
=  
\left( - \hat{\hat{L}} +  L^{+} \right) \hat{\rho} / L^{-}  
= 
\left( - \left[  \hat{H}, \hat{\rho} \right] +  L^{+} \hat{\rho} \right) / L^{-}  \notag \\
= 
&
\left\{ \left[ ( - \hat{H} ), \hat{\rho} \right] + \hat L^{+} \hat{\rho} \right\} / L^{-} \notag \\
= 
&
\left\{ \left[ ( - \hat{H} + \hat{I} L^{+} ), \hat{\rho} \right] + L^{+} \hat{\rho} \right\} / L^{-}  \notag \\
= 
&\left[ ( - \hat{H} + \hat{I} L^{+} )/ L^{-}, \hat{\rho} \right] + L^{+} \hat{\rho}  / L^{-} 
\end{align}
Here,
 If $ L_{\textrm{max/min}} = H_{\textrm{max/min}} $ , then
$ L^{-}=R, L^{+}=H_{\textrm{avr}} $ with 
$ R \equiv (H_{\textrm{max}}-H_{\textrm{min}}) / 2  $ and 
$ H_{avr} \equiv (H_{\textrm{max}}+H_{\textrm{min}}) / 2 $.

%
%
%

\begin{table}[b]
\begin{center}%
\begin{tabular}{cccccc}
\hline
\hline
   $Ng=121$  &\quad $r_{\textrm{pair}}$  &\quad $N_{\textrm{st}}^{\textrm{pair}}$ &\quad CPUt  &\quad  ratio  \\
\hline
             &\quad 8.8  &\quad 1820  &\quad 1.5  &\quad  0.11    \\
             &\quad 9.8  &\quad 2316  &\quad 1.6  &\quad  0.11    \\
             &\quad 10.8  &\quad 2780  &\quad 1.8  &\quad  0.13   \\
             &\quad 15.8  &\quad 5068  &\quad 2.4  &\quad  0.18   \\
\hline
\hline
   $Ng=441$  &\quad $r_{\textrm{pair}}$  &\quad $N_{\textrm{st}}^{\textrm{pair}}$ &\quad CPUt  &\quad  ratio  \\
\hline
             &\quad 8.8  &\quad 7028  &\quad 45  &\quad  0.067   \\
             &\quad 9.8  &\quad 8868  &\quad 45  &\quad  0.067   \\
             &\quad 10.8  &\quad 10998  &\quad 48  &\quad  0.072   \\
             &\quad 15.8  &\quad 20156  &\quad 58  &\quad  0.087   \\
\hline
\hline
   $Ng=961$  &\quad $r_{\textrm{pair}}$  &\quad $N_{\textrm{st}}^{\textrm{pair}}$ &\quad CPUt  &\quad  ratio  \\
\hline
             &\quad 8.8  &\quad 15164  &\quad 906  &\quad  0.047   \\
             &\quad 9.8  &\quad 19676  &\quad 920  &\quad  0.048   \\
             &\quad 10.8  &\quad 24052  &\quad 933  &\quad  0.049   \\
             &\quad 15.8  &\quad 45076  &\quad 1003  &\quad  0.053   \\
\hline
\end{tabular}
\end{center}
\caption{Speedup of dynamics calculation using sparsity of interaction. 
We presented here comparisons 
with respect to pair threshold distance $r_{\textrm{pair}}$ and system size $N_{\textrm{g}}$. 
Moderately disordered case with $f=0.5$ is used.
'CPUt' means a CPU time took in calculations and 
'ratio' shows a ratio of CPU time in case with sparse Hamiltonian to 
that with full size Hamiltonian. 
}
\label{speedup}
\end{table}

\clearpage


\end{document}